\documentclass[aps,prb,twocolumn]{revtex4}   % style for Physical Review B and AJP are similar

\usepackage{amsmath}    % need for subequations
\usepackage{amssymb}
\usepackage{graphicx}   % for figures
\usepackage{times}

\begin{document}

\title{High-energy long-lived resonance of electrons in fractal-like semiconductor heterostructures}
%Lines break automatically or can be forced with \\
\author{Seng Pei Liew}
% \altaffiliation[Also at ]{home.}  %  optional
 \affiliation{Department of Physics, University of Tokyo, Hongo 7-3-1, Bunkyo, Tokyo 113-0033, Japan}
 \email{s101569@mail.ecc.u-tokyo.ac.jp}   %optional
\author{Naomichi Hatano}
\affiliation{Institute of Industrial Science, University of Tokyo, Komaba 4-6-1, Meguro, Tokyo 153-8505, Japan}
\email{hatano@iis.u-tokyo.ac.jp}
\date{\today}

\begin{abstract}
A fractal-like alignment of quantum wells is shown to accommodate resonant states with long lifetimes.
For the parameters of the semiconductor heterostructure GaAs/Al$_{0.4}$Ga$_{0.6}$As with the well depth 300meV, 
a resonant state of the energy as high as 44meV with the lifetime as long as 2.8$\mu$s is shown to be achievable.
\end{abstract}
\maketitle

The propagation of an electromagnetic wave in a fractal object has attracted attention since Takeda \textit{et al}.\cite{Takeda04} experimentally detected a strong resonance of microwave in a Menger sponge of dielectric substance.
They observed a sharp resonance dip of a $Q$ factor as large as 610 at the frequency 12.8GHz in a third-generation Menger sponge of length 27mm.
Subsequently, one of the present authors\cite{Hatano05} as well as Esaki \textit{et al}.\cite{Esaki09} theoretically studied the propagation of an electromagnetic wave in Cantor-set-like structures.
They discovered fractal-like behavior of the transmission and reflection coefficients.
The former study particularly found resonant states of lifetimes as long as of the order of 1ms in the fourth-generation Cantor set of length 10cm and demonstrated strong localization of the wave amplitude inside the Cantor set at such a resonant state.

\begin{figure*}
\includegraphics[width=0.75\textwidth]{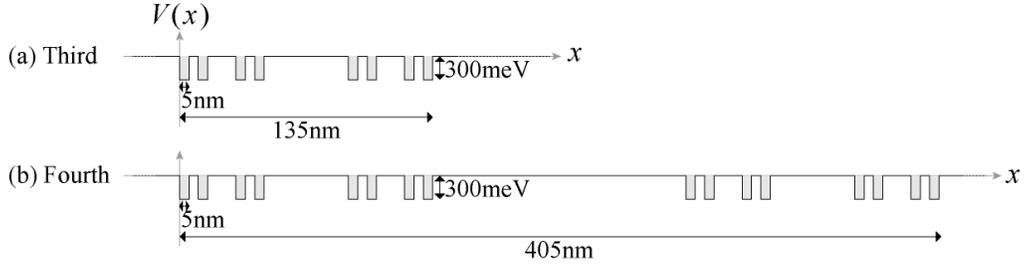}
\caption{(a) The third-generation Cantor set and (b) the fourth-generation Cantor set of quantum wells.
We chose the parameter values that may be easily realized in the semiconductor heterojunction GaAs/Al$_{0.4}$Ga$_{0.6}$As.
The quantum wells of GaAs are embedded in the background Al$_{0.4}$Ga$_{0.6}$As.}
\label{fig1}
\end{figure*}

Due to the similarities between the electromagnetic wave and the quantum-mechanical wave function, it is natural to anticipate similar phenomena within the framework of quantum mechanics.
In the present Report, we will theoretically show that resonant states with long lifetimes indeed exist for a quantum-mechanical particle propagating in Cantor-set like structures (Fig.~\ref{fig1}) and suggest that such long-lived resonant states should be experimentally detected in properly fabricated semiconductor heterostructures.

It is important to find in a semiconductor heterostructure a resonant state with a high eigenenergy but with a long lifetime.
Within its resonance lifetime, a resonant state effectively behaves as a bound state;
then we may be able to use a long-lived resonant state as a highly excited bound state in laser emission and many other applications.
In order to demonstrate that long-lived resonant states are within our reach in a semiconductor heterostructure, we will choose parameter values that may be easily realized in quantum wells of the GaAs/AlGaAs semiconductor heterojunction.
A layer of GaAs in the background of Al$_x$Ga$_{1-x}$As forms a potential well in one dimension.
%S. Adachi, JAP58,R1(1985)
%Band gap: GaAs 1.424ev, AlAs 2.168eV, 
%Al0.4Ga0.6As, the potential well 1.247*0.4*0.6eV=0.299eV
%0.6 of the Band gap for the conduction electron, 0.4 for the valence electron
%meff=0.067+0.083x for AlxGa1-xAs
The depth of the potential well for GaAs in the background of Al$_{0.4}$Ga$_{0.6}$As is considered to be $V=300$meV.\cite{Adachi85}
The electron mass should be the effective mass of an electron in respective region. 
We will use the effective mass\cite{Adachi85} of $m_1=(0.067+0.083\times 0.4)m_0=0.10m_0$ for Al$_{0.4}$Ga$_{0.6}$As outside the potential well and $m_2=0.067m_0$ for GaAs inside the potential well, where $m_0$ is the bare mass of an electron.

We will arrange the quantum wells of width $\ell=5$nm and depth $V=300$meV so that they may constitute finite-generation Cantor sets.
(The Cantor set is operationally built by filling out middle third parts of the potential wells in each generation when we start from one potential well of length $L$ in the zeroth generation.
Strictly speaking, it becomes a fractal object in the infinite-generation limit, but we will use finite-generation Cantor sets here as fractal-like objects.)
The third-generation and fourth-generation Cantor sets come to $L=\ell\times3^3=135$nm and $L=\ell\times3^4=405$nm in total length, respectively, as shown in Fig.~\ref{fig1}.
The fifth-generation Cantor set is of length $L=\ell\times 3^5=1215$nm (not shown in Fig.~\ref{fig1}).

In the present article, for demonstration purposes, we will ignore the effects of impurity scatterings, phonon scatterings, and fluctuations of the depth and the width of the quantum wells.
In practice, these effects may weaken the resonances that we will discuss hereafter.

\begin{figure}
\centering
\includegraphics[width=0.8\columnwidth]{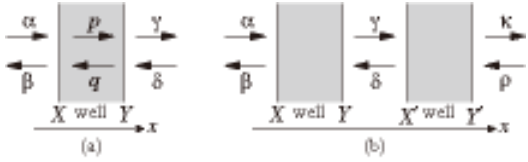}
\caption{(a) Wave amplitudes straddling a potential well (the shaded area).
(b) Wave amplitudes straddling multiple wells (the shaded areas) of the first-generation Cantor set.}
\label{fig2}
\end{figure}

We begin by solving the Schr\"{o}dinger equation for the wave function $\psi(x)$.
We set the wave function of the electron outside the well (in the region of Al$_{0.4}$Ga$_{0.6}$As) as
\begin{align}\label{eq10}
\psi (x) = \alpha e^{+ik_1x}+\beta e^{-ik_1 x},
\end{align}
where $k_1=\sqrt{2m_1E/\hbar^2}$ with Greek letters denoting the wave amplitudes,
while we set the wave function of the electron in the well (in the region of GaAs) as
\begin{align}\label{eq20}
\psi (x) = p \mathrm{e}^{+ ik_2x}+q e^{-ik_2 x},
\end{align}
where  $k_2=\sqrt{2m_2(E+V)/\hbar^2}$ with roman letters denoting the wave amplitudes (Fig.~\ref{fig2}).
Note that we use the different effective masses for the two regions.

We now use the transfer-matrix method to compute the wave amplitudes in the Cantor set.
The standard boundary conditions at $x=X$ in Fig.~\ref{fig2}(a) between the background region and the well region lead to 
\begin{align}\label{eq30}
&\begin{pmatrix}
\mathrm{e}^{ik_1X}&\mathrm{e}^{-ik_1X}\\
ik_1\mathrm{e}^{ik_1X}&-ik_1\mathrm{e}^{ik_1X}\\
\end{pmatrix}
\begin{pmatrix}
\alpha\\
\beta\\
\end{pmatrix}
\nonumber\\
&\qquad=
\begin{pmatrix}
\mathrm{e}^{ik_2X}&\mathrm{e}^{-ik_2X}\\
ik_2\mathrm{e}^{ik_2X}&-ik_2\mathrm{e}^{ik_2X}\\
\end{pmatrix}
\begin{pmatrix}
p\\
q\\
\end{pmatrix}.
\end{align}
We obtain a similar relation between the wave amplitudes $(p,q)$ and $(\gamma, \delta)$ for the boundary at $x=Y$ in Fig.~\ref{fig2}(a).
We can cast the above two relations into the form of a transfer matrix that relates $(\alpha,\beta)$ to $(\gamma,\delta)$ as follows:\cite{Hatano05}
\begin{align}\label{eq40}
e^{ik_1Y\sigma_z}
\begin{pmatrix}
\gamma \\
i\delta
\end{pmatrix}
=T^{(0)}(k_1\ell)
e^{ik_1X\sigma_z}
\begin{pmatrix}
\alpha \\
i\beta
\end{pmatrix}.
\end{align}
The transfer matrix $T^{(0)}$ in Eq.~\eqref{eq40} is the matrix that transfers the wave amplitudes over a potential well of width $\ell=Y-X$ and is given by the expression independent of $X$ and $Y$:
\begin{align}\label{eq50}
T^{(0)}(\zeta)=e^{i\zeta n(\sigma_z\cosh\varphi+\sigma_x\sinh\varphi)}
\end{align}
with
%\begin{align}\label{eq60}
$n=k_2/k_1=\exp(\varphi)$.
%\end{align}
In above, $\sigma_\mu$ are the Pauli matrices.
On the other hand, the matrix that transfers the wave amplitudes over a background region of width $\ell=X'-Y$ in Fig.~\ref{fig2}(b) is given by $T_1(k_1\ell)$ with
\begin{align}\label{eq70}
T_1(\zeta)=e^{i\zeta\sigma_z}.
\end{align}

The transfer matrix that transfers the wave amplitudes over the first-generation Cantor set (Fig.~\ref{fig2}(b)) is then given by
\begin{align}\label{eq80}
T^{(1)}(3k_1\ell)=T^{(0)}(k_1\ell)
T_1(k_1\ell)
T^{(0)}(k_1\ell).
\end{align}
The transfer matrix for higher-generation Cantor sets is recursively given by
\begin{align}\label{eq90}
T^{(\nu+1)}(3\zeta)=T^{(\nu)}\left(\zeta\right)
T_1\left(\zeta\right)
T^{(\nu)}\left(\zeta\right).
\end{align}
We can thereby arrive at the transfer matrix that transfers the wave amplitudes over the entire Cantor set of the $\nu$th generation in the form
\begin{align}\label{eq100}
e^{ik_1L\sigma_z}
\begin{pmatrix}
A_{\mathrm{Rout}}\\
A_{\mathrm{Rin}}
\end{pmatrix}
=
T^{(\nu)}(k_1L)
\begin{pmatrix}
A_{\mathrm{Lin}}\\
A_{\mathrm{Lout}}
\end{pmatrix},
\end{align}
where the coefficients $A_{\mathrm{Lin}}$ and $A_{\mathrm{Lout}}$ denote the amplitudes of the incoming and outgoing waves on the left of the Cantor set, respectively, while $A_{\mathrm{Rin}}$ and $A_{\mathrm{Rin}}$ denote those on the right, respectively.
%(Here and hereafter we drop the phase factors that were present in Eq.~\eqref{eq40} for brevity's sake.)

We now introduce the resonance condition.
We can define a quantum-mechanical resonant state as an eigenstate of the Schr\"{o}dinger equation under the boundary condition of outgoing waves only,\cite{Gamow28,Siegert39,Peierls59,Landau77,Hatano08} which is often called the Siegert boundary condition.
Such states are not realized for real values of the energy because they do not conserve the particle number in the standard sense.
(See an argument in Ref.~\cite{Hatano08} for the particle-number conservation in a generalized sense.)
Their eigenenergies are generally complex.
However, they manifest themselves in the form of resonance peaks in the energy dependence of the transmission and reflection coefficients for real values of the energy.
Particularly when the eigenenergy is close to the real energy axis, the $Q$ factor of the corresponding resonance peak, which is inversely proportional to the imaginary part of the complex eigenenergy, becomes prominent.
At the same time, the resonance lifetime, which is also inversely proportional to the imaginary part of the complex eigenenergy, becomes large and hence the resonant state is more stabilized.
As stated above, our aim here is to find highly excited and long-lived resonant states.
In other words, we look for a resonant state with a large real part and a small imaginary part of the eigenenergy.
%Within the resonance lifetime, the scattering state with the real part of the complex eigenenergy may effectively behave as a bound state.
%It is therefore important to find in a semiconductor structure a resonant state with a high energy but with a long lifetime;
%then we may be able to use it as a highly excited bound state in laser emission and other applications.

Some readers might wonder why resonance occurs in a structure with attractive potentials only such as the present system.
Indeed, resonance is typically considered in a structure with multiple potential walls.
In fact, resonance can occur even in the present system because the wave is reflected back and forth at the potential discontinuities.

\begin{figure}
\centering
\includegraphics[width=0.8\columnwidth]{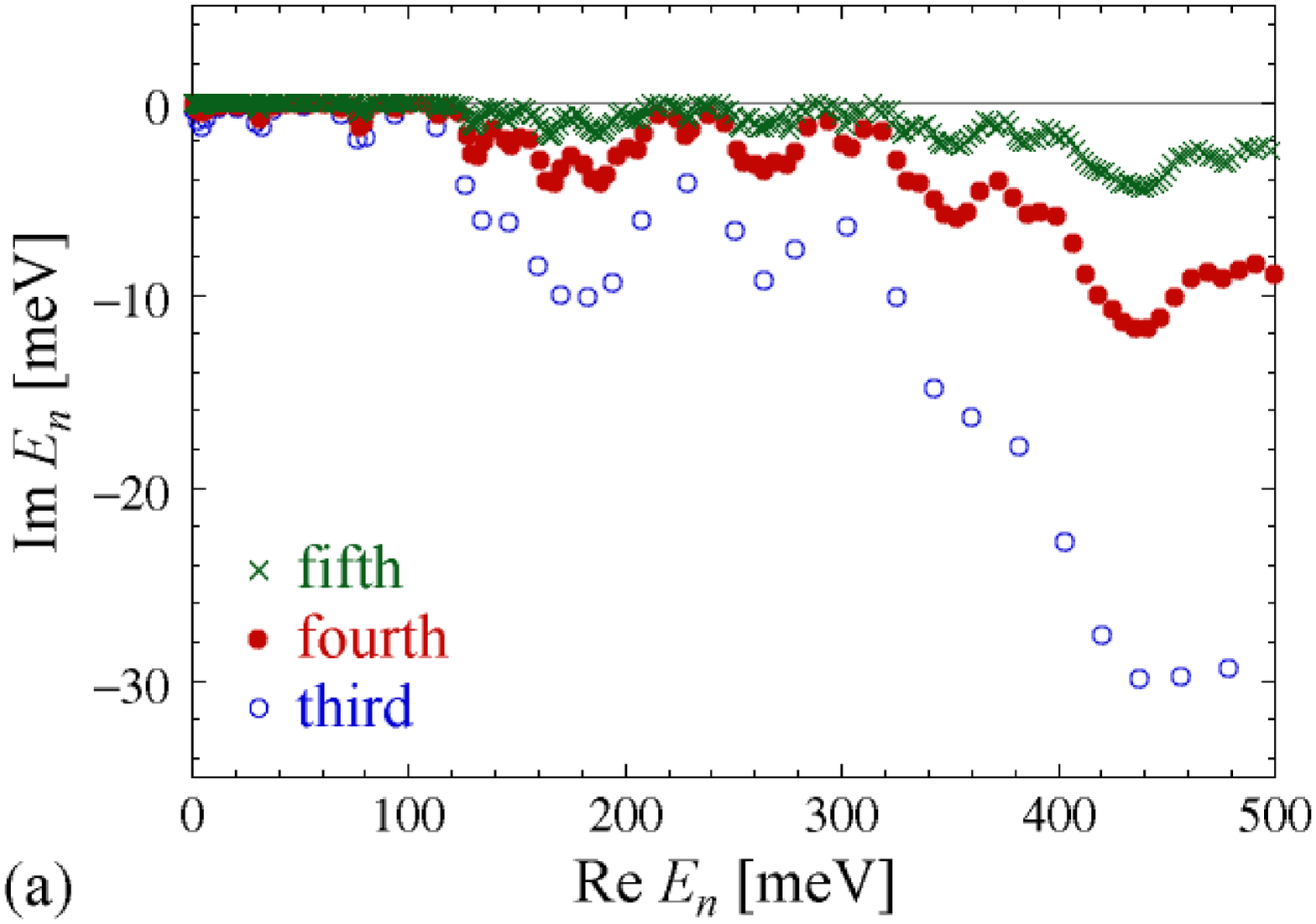}\\
\vspace{\baselineskip}
\includegraphics[width=0.8\columnwidth]{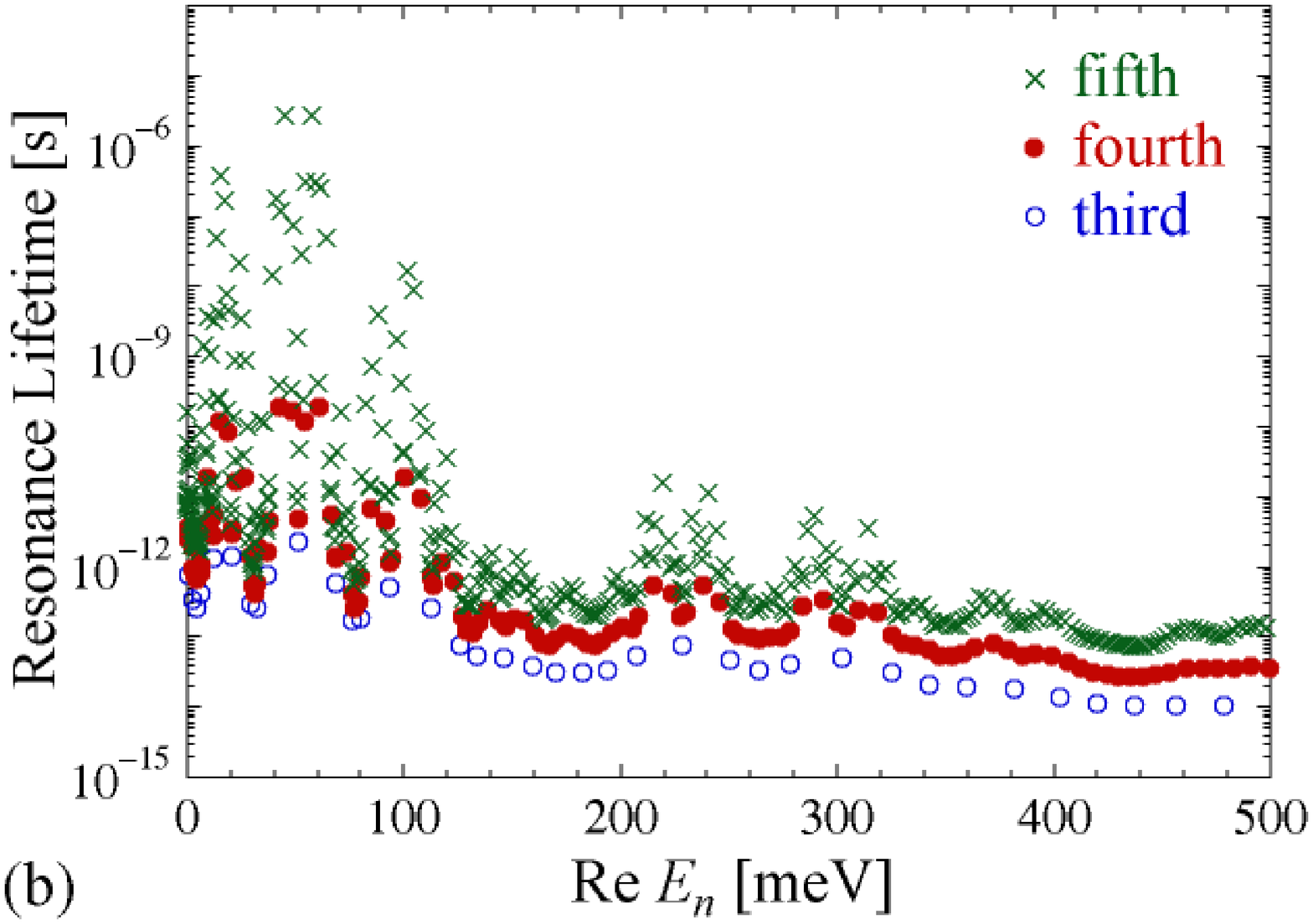}
\caption{(color online) (a) The resonance eigenvalues (the solutions of the resonance equation~\eqref{eq110}) plotted on the complex energy plane.
(b) The resonance lifetime of each resonant state plotted against the real part of its resonance eigenvalue.
In both panels, the open (blue) circles, the solid (red) circles and the (green) crosses indicate the eigenvalues for the third-, fourth- and fifth-generation Cantor sets, respectively.}
\label{fig3}
\end{figure}

The Siegert boundary condition of out-going waves only is given by putting $A_{\mathrm{Lin}}=A_{\mathrm{Rin}}=0$ in Eq.~\eqref{eq100}. This leads to the resonance condition in the form
\begin{align}\label{eq110}
T^{(\nu)}_{22}(\zeta_n)=0.
\end{align}
This determines the complex eigenvalues of $k_1=\zeta_n/L$ and then the complex eigenenergies $E=\hbar^2{k_1}^2/(2m_1)=\hbar^2{\zeta_n}^2/(2mL^2)$.
(The origin of the energy is set at the Fermi energy of the background Al$_{0.4}$Ga$_{0.6}$As.
The bottom of the quantum well is at $-300$meV.)
Figure~\ref{fig3}(a) shows the solutions of the resonance equation~\eqref{eq110} for the third-, fourth- and fifth-generation Cantor sets.
(We numerically solved the resonance equation by the Newton-Lapson method.)
The imaginary parts of the resonance eigenvalues are all negative as they should be.

The resonance lifetime is given by
\begin{align}
\tau_n=\frac{\hbar}{\left|\mathop{\mathrm{Im}}E_n\right|};
\end{align}
then the eigenfunction decays as
\begin{align}
\left|\Psi_n(x,t)\right|^2=\left|\psi_n(x)\right|^2 e^{-t/\tau_n}.
\end{align}
We plotted the resonance lifetime of each resonant state in Fig.~\ref{fig3}(b).
The lifetime becomes longer in some regions.
The most prominent ones are listed in Tab.~\ref{tab1}.
\begin{table}[t]
\caption{Resonant states with the longest lifetimes:
(a) the states with $\tau_n>1$ps for the third-generation Cantor set;
(b) the states with $\tau_n>0.01$ns for the fourth-generation Cantor set;
(c) the states with $\tau_n>0.01\mu$s for the fifth-generation Cantor set.
The symbol $\ast$ indicate the state with the longest lifetime in each generation.
Note that the units are different for different Cantor sets in the second and third columns.}
\label{tab1}
\vspace{\baselineskip}
\begin{tabular}{r@{.}l r@{.}l r@{.}l}
\hline
\multicolumn{6}{l}{(a) Third-generation Cantor set} \\
\hline
\multicolumn{2}{c}{$\mathop{\mathrm{Re}}E_n$[meV]} & 
\multicolumn{2}{c}{$\mathop{\mathrm{Im}}E_n$[meV]} & 
\multicolumn{2}{c}{Lifetime $\tau_n$[ps]} \\
\hline\hline
$11$&$5481$ & $-0$&$243062$ & $1$&$34997$ \\
$20$&$5725$ & $-0$&$225531$ & $1$&$45490$ \\
$51$&$2064$ & $-0$&$135748$ & $2$&$41716\quad\ast$\\
\hline
\multicolumn{6}{l}{}\\
\hline
\multicolumn{6}{l}{(b) Fourth-generation Cantor set} \\
\hline
\multicolumn{2}{c}{$\mathop{\mathrm{Re}}E_n$[meV]} & 
\multicolumn{2}{c}{$\mathop{\mathrm{Im}}E_n$[$\mu$eV]} & 
\multicolumn{2}{c}{Lifetime $\tau_n$[ns]} \\
\hline\hline
$ 8$&$83250$ & $-16$&$2906$  & $0$&$0201420$ \\
$14$&$5938$  & $ -2$&$74546$ & $0$&$119515$ \\
$17$&$9402$  & $ -3$&$80652$ & $0$&$0862007$ \\
$21$&$9982$  & $-19$&$7782$  & $0$&$0165903$ \\
$25$&$7567$  & $-17$&$1651$  & $0$&$0191159$ \\
$42$&$2383$  & $ -1$&$64424$ & $0$&$199560$ \\
$47$&$8313$  & $ -1$&$88180$ & $0$&$174367$ \\
$54$&$0592$  & $ -2$&$64737$ & $0$&$123943$ \\
$60$&$3342$  & $ -1$&$60856$ & $0$&$203987\quad\ast$ \\
$99$&$5199$  & $-16$&$3889$  & $0$&$0200212$ \\
\hline
\multicolumn{6}{l}{}\\
\hline
\multicolumn{6}{l}{(c) Fifth-generation Cantor set} \\
\hline
\multicolumn{2}{c}{$\mathop{\mathrm{Re}}E_n$[meV]} & 
\multicolumn{2}{c}{$\mathop{\mathrm{Im}}E_n$[neV]} & 
\multicolumn{2}{c}{Lifetime $\tau_n$[$\mu$s]} \\
\hline\hline
$ 13$&$06101$ & $ -6$&$58518 $ & $0$&$0498278$ \\
$ 15$&$08510$ & $ -0$&$847746$ & $0$&$387056 $ \\
$ 17$&$00385$ & $ -1$&$91049 $ & $0$&$171749 $ \\
$ 23$&$07059$ & $-15$&$4223  $ & $0$&$0212760$ \\
$ 39$&$00330$ & $-22$&$3726  $ & $0$&$0146664$ \\
$ 40$&$08315$ & $ -1$&$82814 $ & $0$&$179485 $ \\
$ 42$&$07149$ & $ -2$&$75198 $ & $0$&$119232 $ \\
$ 44$&$06494$ & $ -0$&$116810$ & $2$&$80906  \quad\quad\ast$ \\
$ 48$&$06551$ & $ -4$&$40451 $ & $0$&$0744975$ \\
$ 52$&$09428$ & $-11$&$4234  $ & $0$&$0287238$ \\
$ 55$&$00840$ & $ -1$&$04331 $ & $0$&$314503 $ \\
$ 57$&$02796$ & $ -0$&$122624$ & $2$&$67587  $ \\
$ 59$&$05157$ & $ -1$&$05117 $ & $0$&$312152 $ \\
$ 61$&$07859$ & $ -1$&$29655 $ & $0$&$253075 $ \\
$ 64$&$00625$ & $ -6$&$78678 $ & $0$&$0483477$ \\
$102$&$1005 $ & $-19$&$6566  $ & $0$&$0166928$ \\
\hline
\end{tabular}
\end{table}
%(We listed the values of $|E_n|$ as well, because a recent study\cite{Klaiman10} found that the peak location of a resonance is given by $|E_n|$ rather than $\mathop{\mathrm{Re}}E_n$.)
(For references, we also list the bound states in Tab.~\ref{tab2}.
Note again that the bottom of the quantum well is at $-300$meV.)
\begin{table}[t]
\caption{The bound-state eigenenergies $E_n$[meV]:
(a) for the third-generation Cantor set;
(b) for the fourth-generation Cantor set;
(c) for the fifth-generation Cantor set.}
\label{tab2}
\vspace{\baselineskip}
\begin{tabular}{r@{.}l r@{.}l r@{.}l}
\hline
\multicolumn{2}{l}{(a) Third} &
\multicolumn{2}{l}{(b) Fourth} &
\multicolumn{2}{l}{(c) Fifth} \\
\hline
$-201$&$9843 $       & $-201$&$9844    $    & $-201$&$9844    $ \\
$-201$&$9835 $       & $-201$&$9835    $    & $-201$&$9839    $ \\
$-199$&$3396 $       & $-199$&$3396    $    & $-201$&$9835    $ \\
$-199$&$3386 $       & $-199$&$3386    $    & $-199$&$3396    $ \\
$- 15$&$91427$       & $- 15$&$91427   $    & $-199$&$3391    $ \\
$- 15$&$91325$       & $- 15$&$91325   $    & $-199$&$3386    $ \\
$- 14$&$81840$       & $- 15$&$91427   $    & $- 15$&$91427   $ \\
$- 14$&$81676$       & $- 15$&$91325   $    & $- 15$&$91325   $ \\
\multicolumn{2}{c}{} & $-  0$&$01114612$    & $- 14$&$81840   $ \\
\multicolumn{2}{c}{} & \multicolumn{2}{c}{} & $- 14$&$81676   $ \\
\multicolumn{2}{c}{} & \multicolumn{2}{c}{} & $-  0$&$01192419$ \\
\multicolumn{2}{c}{} & \multicolumn{2}{c}{} & $-  0$&$01024101$ \\
\hline
\end{tabular}
\end{table}
For the fourth-generation Cantor set of length $L=405$nm, the longest lifetime was $0.2$ns found at $60$meV.
A shorter one of $0.02$ns was found at a higher energy of $100$meV.
For the fifth-generation Cantor set of length $L=1215$nm, the longest lifetime was $2.8\mu$s found at $44$meV.
A shorter one of $0.02\mu$s was found at a higher energy of $102$meV.
We can see that the lifetime grows very rapidly as we go to higher generations.

\begin{figure}[t]
\hspace{-1.5mm}
\includegraphics[height=0.6\columnwidth]{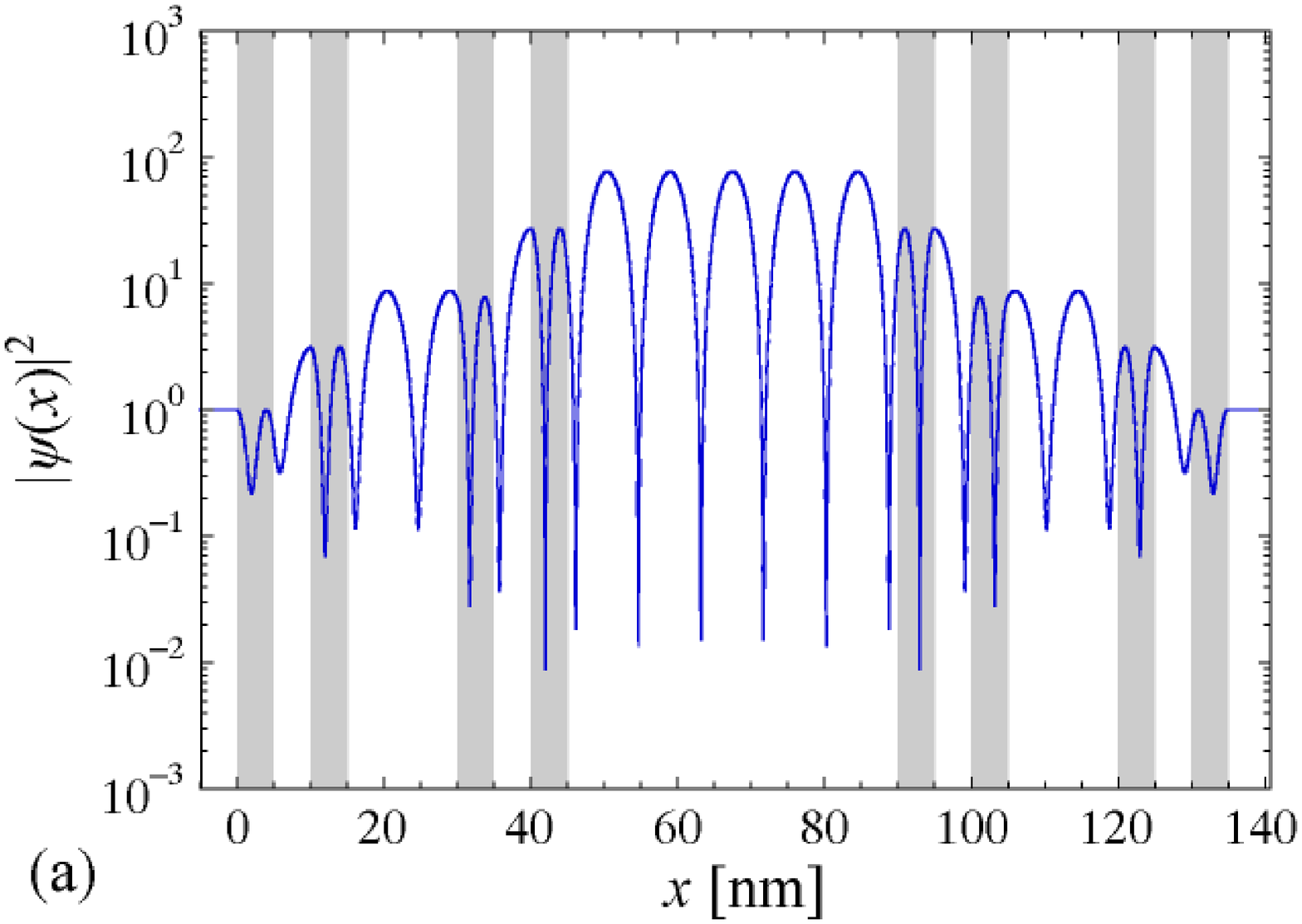}\\
\vspace{\baselineskip}
\hspace{-3mm}
\includegraphics[height=0.6\columnwidth]{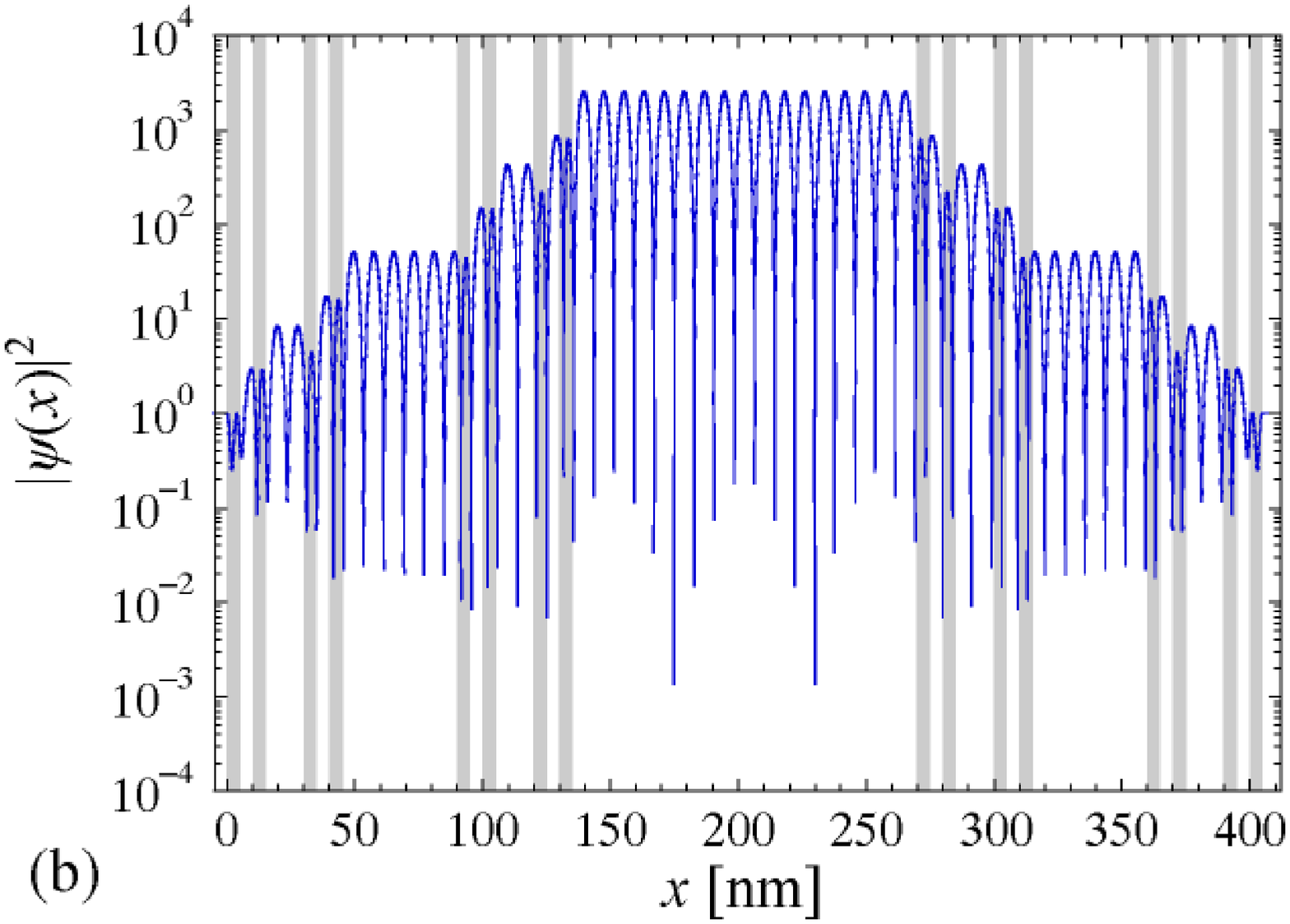}\\
\vspace{\baselineskip}
\hspace{-1.5mm}
\includegraphics[height=0.6\columnwidth]{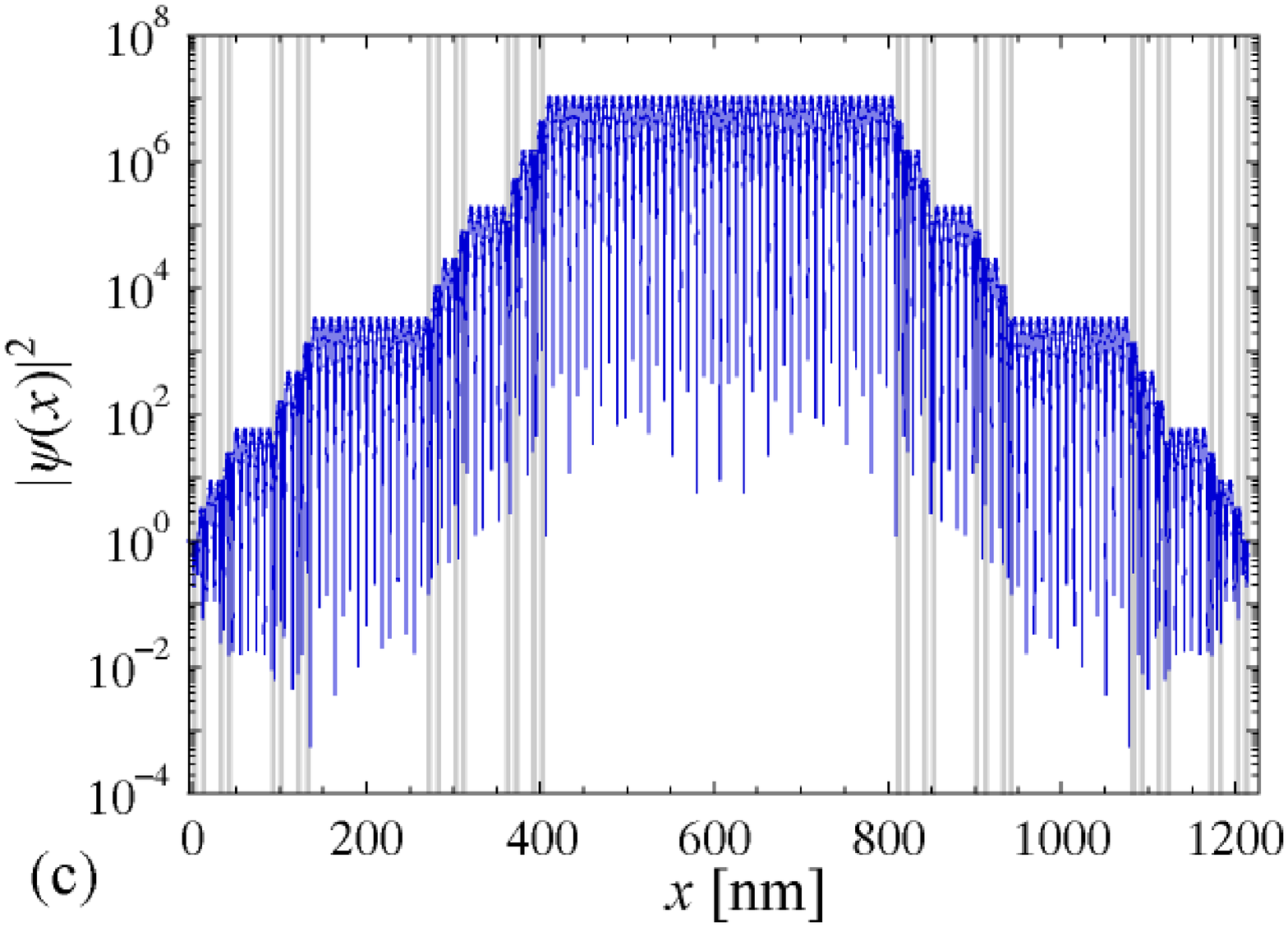}\\
\caption{(color online) Semi-logarithmic plots of the wave amplitudes $|\psi_k(x)|^2$ of the scattering states with (a) $k=0.368336$[1/nm] for the third-generation Cantor set, (b) $k=0.399816$[1/nm] for the fourth-generation Cantor set, and (c) $k=0.343943$[1/nm] for the fifth-generation Cantor set.
Each wave number is equal to the real part of the eigen-wave-number of the resonant state with the longest lifetime.
The shaded columns indicate the regions of the quantum wells of GaAs.}
\label{fig4}
\end{figure}

Figure~\ref{fig4} shows the amplitude of the wave function of a scattering state that has the wave number equal to the real part of the eigen-wave-number of the resonant state with the longest lifetime in each generation (marked by the symbol $\ast$ in Tab.~I).
Note that the wave numbers and the energies of these scattering states are all real;
hence these states are physically possible to make. 
We can see strong localization of the wave amplitude in the central Al$_{0.4}$Ga$_{0.6}$As portion of the heterostructure.
This demonstrates that the resonance with a long lifetime can effectively bind the electron inside the system.

To summarize, we found long-lived resonant states for a series of quantum wells aligned in the form of a Cantor set, a fractal object.
We numerically calculated the lifetimes for the parameter values that correspond to quantum wells of GaAs embedded in the background of Al$_{0.4}$Ga$_{0.6}$As.
The lifetime is as long as $\sim0.1$ns for the fourth-generation Cantor set and $\sim1\mu$s for the fifth-generation Cantor set;
they are found at about $50$meV for the potential wells of depth $-300$meV.
Relatively long-lived resonant states are found even around $100$meV.
There are also clusters of resonant states of lifetimes longer than others around $\sim220$meV and $\sim300$meV.
The wave amplitude of the corresponding scattering state is strongly localized in the central part of the system, which shows effective binding of an electron.

As mentioned above, we here ignored the effects of impurity scatterings, phonon scatterings, and fluctuations of the depth and the width of the quantum wells.
Although these effects may, to some extent, shorten the resonance lifetimes that we found in the ideal situation, the resonances are still worth pursuing experimentally for their possible abundant applications.

We are grateful to Prof. S. Iwamoto's comments from an experimental point of view.

\end{document}